\documentclass{jfm}
                                                                                             
\ifCUPmtlplainloaded \else
  \checkfont{eurm10}
  \iffontfound
    \IfFileExists{upmath.sty}
      {\typeout{^^JFound AMS Euler Roman fonts on the system,
                   using the 'upmath' package.^^J}%
       \usepackage{upmath}}
      {\typeout{^^JFound AMS Euler Roman fonts on the system, but you
                   dont seem to have the}%
       \typeout{'upmath' package installed. JFM.cls can take advantage
                 of these fonts,^^Jif you use 'upmath' package.^^J}%
      }
  \else
  \fi
\fi
\ifCUPmtlplainloaded \else
  \checkfont{msam10}
  \iffontfound
    \IfFileExists{amssymb.sty}
      {\typeout{^^JFound AMS Symbol fonts on the system, using the
                'amssymb' package.^^J}%
       \usepackage{amssymb}%

      }{}
  \fi
\fi
                                                                                             
\ifCUPmtlplainloaded \else
  \IfFileExists{amsbsy.sty}
    {\typeout{^^JFound the 'amsbsy' package on the system, using it.^^J}%
     \usepackage{amsbsy}}
    {\providecommand\boldsymbol[1]{\mbox{\boldmath $##1$}}}
\fi

\newsavebox{\astrutbox}
\sbox{\astrutbox}{\rule[-5pt]{0pt}{20pt}}

\newcommand\etal{\mbox{\textit{et al.}}}

\usepackage{times}
\usepackage{epsfig}
\input epsf

\newcommand{\pl}{\partial}

\title[New instability mode in a plane thermal plume]{Effects of Prandtl number and a new 
instability mode in a plane thermal plume}
\author[R. Lakkaraju and M. Alam ]%
{
\textsc{R. LAKKARAJU}
\textsc{and}
\textsc{MEHEBOOB ALAM}\footnote{Author to whom correspondence should be addressed:
Email: meheboob@jncasr.ac.in \\
Published in {\it J. Fluid Mech.}, vol. 592, p. 221-231 (2007)
}
}
\affiliation{
 Engineering Mechanics Unit, Jawaharlal Nehru Center for Advanced
 Scientific Research\\
 Jakkur P.O., Bangalore 560064, India
}
\pubyear{2007}
\volume{592}
\pagerange{221--231}
\date{1 July 2007 and in revised form 10 September 2007}
\begin{document}

\maketitle

\begin{abstract}
The effect of Prandtl number on the linear stability of a plane 
thermal plume is analyzed under quasi-parallel approximation.
At large Prandtl numbers ($Pr>100$), we found that  
there is an additional unstable loop whose size increases with increasing $Pr$. 
The origin of this new instability mode is shown to be tied to the 
coupling of the momentum and thermal perturbation equations.   
Analyses of the perturbation kinetic energy and thermal energy suggest that 
the buoyancy force is the main source of
perturbation energy at high Prandtl numbers that drives this instability. 
\end{abstract}

\section{Introduction}

The classic problem of  natural-convection flow above a horizontal line heat source 
has received considerable attention during the last few decades (Batchelor 1954;
Fujii 1963; Gebhart \etal 1988).
The temperature of the heat source is larger than that of the ambient fluid, 
and the resulting density difference creates a plume that rises up against the gravity.
For steady laminar plumes, the similarity solutions of the pertinent boundary 
layer equations have been published by many researchers
(Fujii 1963; Gebhart, Pera \& Schorr 1970; Riley 1974); 
experimental studies on laminar plane plumes
are in good agreement with similarity solutions (Riley 1974).
Experiments (Pera \& Gebhart 1971) have confirmed  that the laminar plumes
are unstable, and they sway in a plane perpendicular to the axis of the source. 
Pera \& Gebhart (1971) have shown that the initial instability of plane 
plumes to two-dimensional disturbances can be analyzed by 
the linear stability theory and developed the coupled 
Orr-Sommerfeld type equations using a quasi-parallel flow approximation. 
Since Squire's theorem holds for natural convection flows (Gebhart \etal 1988), 
it is sufficient to consider  two-dimensional 
disturbances for the stability analysis of a thermal plume.

Strictly speaking, the thermal plume is a non-parallel flow field,
and the streamwise variations of both the laminar and disturbed flows
should be incorporated in the stability analysis 
(Hieber \& Nash 1975; Wakitani 1985).
From a weakly non-parallel spatial stability analysis (Wakitani 1985),
it has been shown that the critical Grashof number of a plane thermal plume 
is slightly larger than that predicted from
the quasi-parallel theory, even though its precise value depends on the
flow quantity (fluctuating kinetic energy or thermal energy, etc.)
that is being monitored to calculate non-parallel corrections.
It was shown that a lower branch of the neutral stability curve
in the (frequency, Grashof number)-plane exists when the non-parallel corrections
are taken into account. The upper branch of the neutral curve
at moderate-to-large values of Grashof number remains relatively unaffected, however,
with non-parallel corrections.

Two non-dimensional numbers involved in natural convection phenomena are the Grashof
number ($Gr$), the ratio of the buoyancy force and the viscous force, 
and the Prandtl number ($Pr$), the ratio of
the kinematic viscosity and the thermal diffusivity. 
In terms of $Pr$, there are two limiting cases: zero Prandtl number limit 
(e.g. molten metals $Pr\sim 10^{-2}$) 
and infinite Prandtl number limit (e.g. $Pr\sim 10^{3}$ for magmas,
and $Pr\sim 10^{21}$ for Earth's mantle plume).
Geological flows involve fluids with large Prandtl numbers 
(Worster 1986; Lister 1987; Grossman \& Lohse 2000; 
Kaminski \& Jaupert 2003; Majumder, Yuen \& Vincent 2004) 
and are studied in the limit of infinite Prandtl number 
(Wang 2004) for which the inertial
terms in the momentum equations are neglected.

The goal of the present work is to understand
the stability characteristics of high Prandtl number plane thermal plumes.
To the best of our knowledge, all stability analyses 
of plane thermal plumes (Pera \& Gebhart 1971; Hieber \& Nash 1975; Wakitani 1985) 
are confined to that of air ($Pr=0.7$) and water ($Pr=6.7$).
We use the quasi-parallel approximation to analyse the 
linear stability of a thermal plume which is found to be unstable 
for very small Grashof numbers at any Prandtl number. 
At high Prandtl numbers, we find a new instability loop
which is shown to be tied to the coupling of the hydrodynamic and thermal
perturbation equations. An analysis of the perturbation energy unveils
the driving mechanism of this instability.

\section{Governing equations and base flow}

We consider the convective flow generated above a line heat source 
in an otherwise stagnant 
fluid which is maintained at a constant temperature $T_{\infty}$. 
Let the Cartesian coordinate system is ($x, y$), with $x$ being directed along the
flow direction (i.e. against gravity) and $y$  is the transverse
direction, and $u$ and $v$ are the corresponding velocity components along
$x$ and $y$ directions, respectively, and $t$ is the time.
With Boussinesq approximation,
the governing equations for the velocity and the temperature fields are given by
\begin{eqnarray}
{\pl u\over\pl{t}} + u{\pl u\over\pl x} + v{\pl u\over\pl y} &=&
{\nu}{\nabla^2 u} - {1\over\rho}{\pl p\over\pl x} + g{\beta}{(T -T_{\infty})} 
\label{eqn:de2} \\
{\pl v\over\pl{t}} + u{\pl v\over\pl x} + v{\pl v\over\pl y} &=&
\nu \nabla^2 v - {1\over\rho}{\pl p\over\pl y} 
\label{eqn:de3}\\
{\pl T\over\pl{t}} + u{\pl T\over\pl x} + v{\pl T\over\pl y} &=&
{\kappa \nabla^2 T},
\quad\quad 
{\pl u\over\pl x} + {\pl v\over\pl y} \; =\; 0  
\label{eqn:de4}
\end{eqnarray}
Here $\rho$ is the mean density of the fluid, $p$ is the pressure,
$g$ is the acceleration due to gravity;
the thermo-physical properties of the fluid are  
the thermal expansion coefficient $\beta$, 
the kinematic viscosity $\nu$, the thermal conductivity $k$, the specific heat 
at constant pressure $c_p$ and the thermal diffusivity $\kappa = k/\rho c_p$. 
The boundary conditions on velocity and temperature are:
$u = v = 0,{\;}T = {T_s}{\;} {\;}\mbox{at}{\;} x = y = 0{\;}{\;}$ 
and $u = v = T = 0$ at $x^{2} + y^{2} \rightarrow \infty$.

\subsection{Base flow: similarity solution}

The steady laminar base flow is given by the leading-order boundary-layer equations 
(Fujii 1963; Gebhart \etal 1970; Pera \& Gebhart 1971; Riley 1974)
that can be expressed in terms of a stream function:
$u = {\pl\psi}/{\pl y}$, $v = - {\pl\psi}/{\pl x}$.
The resulting partial differential equations (not shown) 
can be transformed into a set of ODEs
in terms of a similarity variable
$\eta = {y}/{\delta}$, with $\delta = {4x}/{G}$,
where
\begin{equation}
   Gr= \frac{g\beta(T_0(x) - T_\infty)x^3}{\nu^2}
\quad\mbox{and} \quad
   G=4\left(\frac{Gr}{4}\right)^{1/4} \quad
\label{eqn:Grashoff1}
\end{equation}
are the local Grashof number and the `modified' Grashof number, respectively,
and $T_0(x)\equiv T(x,y=0)$ is the local centerline temperature.
The non-dimensional stream-function and temperature are defined via
\begin{equation}
       f(\eta) = \frac{\psi}{U_c\delta},
\quad
      h(\eta) = \frac{T-T_\infty}{T_0(x) - T_\infty}
          = \frac{T-T_\infty}{(\nu U_c/g\beta\delta^2)} ,
\label{eqn:nondim1}
\end{equation}
where $U_c=\nu G^2/4x$ is the local convective velocity
and $T_c=(T_0(x) - T_\infty)=\nu U_c/g\beta\delta^2$ is the local excess
temperature at the centerline of the plume.
With the assumption of power-law variation of $T_c$ ($\sim x^{-3/5}$),
the similarity equations can be obtained as 
\begin{equation}
  f^{'''} + {12\over 5}ff^{''} - {4\over 5}{f^{'}}^2 + h = 0, 
\quad 
  h^{''} + {12\over 5}Pr\left(fh\right)^{'} = 0, 
\label{eqn:Base}
\end{equation}
where the prime denotes differentiation with respect to $\eta$,
and the related boundary conditions are (Gebhart \etal 1970):
$f(0) = f^{''}(0) = h^{'}(0) = 0$,  $h(0) = 1$, 
$f^{'}(\infty) \rightarrow 0$,  $h(\infty) \rightarrow 0$. 
These equations have been  solved by using the fourth-order 
Runge-Kutta method with Newton-Raphson correction.
The far-field boundary condition was implemented at $\eta=12$,
and the results were checked by using different values
of $\eta=8, 16, 20, 50$.
Figure 1 shows the velocity and the temperature profiles for a range
of Prandtl numbers.  With increasing $Pr$, the velocity profile flattens 
across the plume width, and the temperature
boundary layer becomes narrower.

%----------------------
\begin{figure}
\begin{center}
\includegraphics[width=5.50cm]{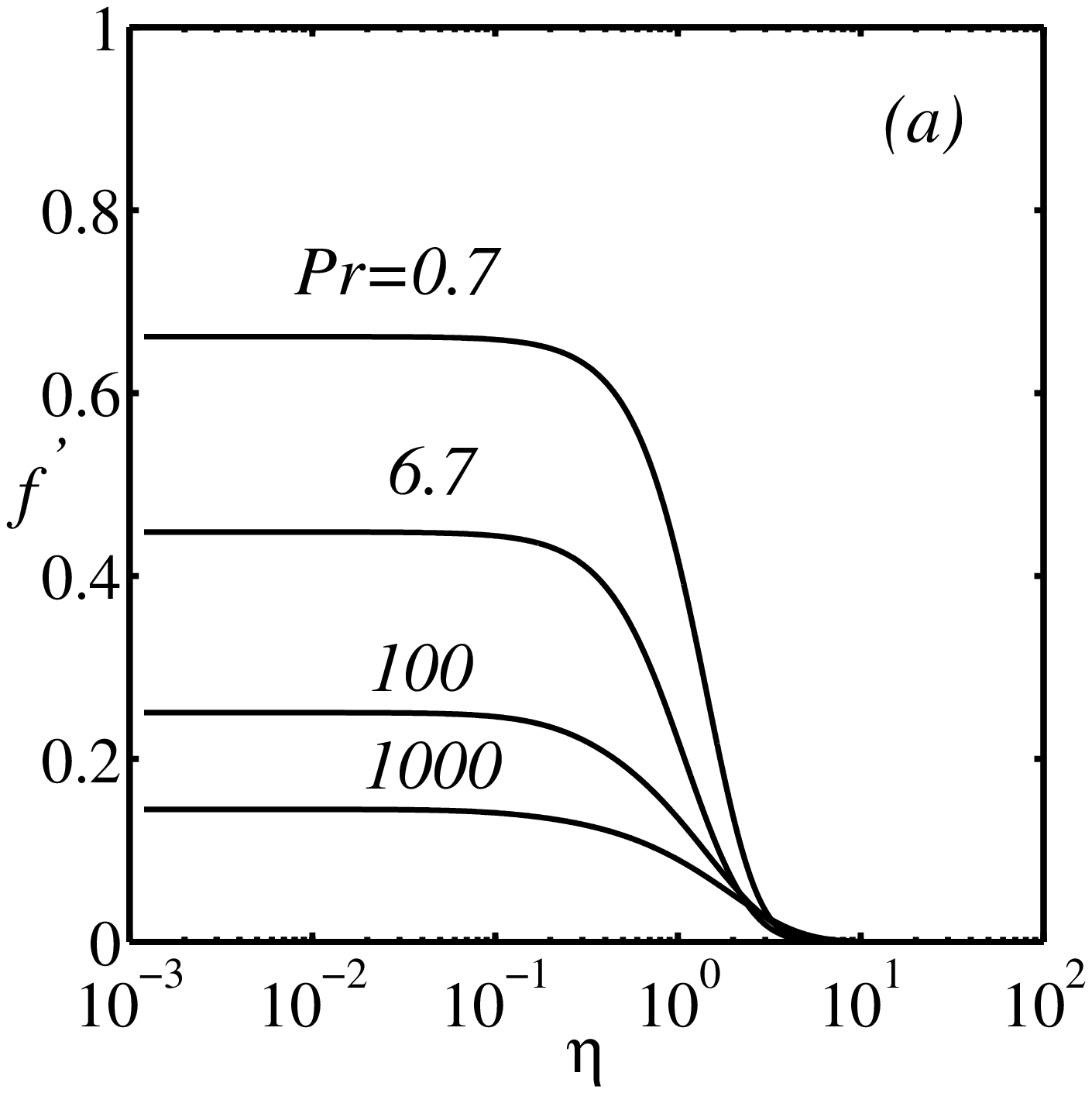} \hspace*{0.8cm}
\includegraphics[width=5.50cm]{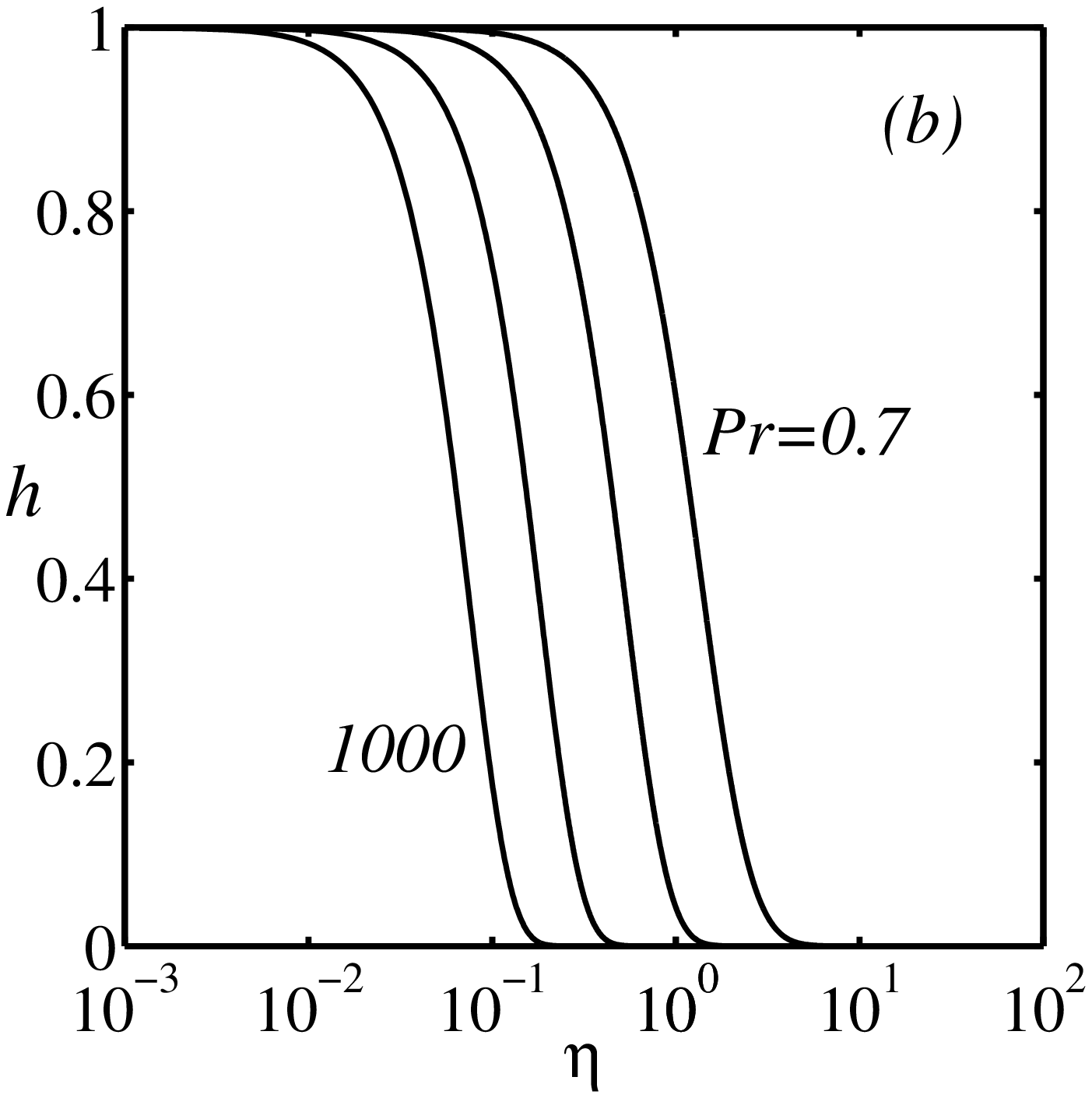}
\end{center}
\caption{
Variations of the base-state ($a$) velocity ($f'$)
and ($b$) temperature ($h$) with Prandtl number.
}
\label{fig:Fig1}
\end{figure}
%----------------------

\section{Linear stability analysis: quasi-parallel approximation}

To analyse the stability of a thermal plume, we decompose each dynamical variable
into a mean part (base flow) and a small-amplitude perturbation:
\begin{eqnarray}
  u(x,y,t) &=& u(x,y) + {\tilde u}(x,y,t), \quad
  v(x,y,t) \;=\; v(x,y) + {\tilde v}(x,y,t) \\
  T(x,y,t) &=& T(x,y) + {\tilde T}(x,y,t), \quad
  {p}(x,y,t) \;=\; p(x,y) + {\tilde p}(x,y,t) 
\end{eqnarray}
with the base flow being taken as that given by the similarity solution 
(\ref{eqn:Base}) over which the perturbation equations are linearized.
The base flow quantities $v$ and the $x$-derivatives of $u$ and $T$ are 
taken as zero in the linearized perturbation equations-- 
this is called the {\it quasi-parallel} approximation. 
The perturbations are assumed to be of the form such that their amplitudes 
depend on the similarity variable $\eta$, as does the base flow.

As in the case of the base flow, it is straightforward 
to show that the perturbation equations 
can be expressed in terms of a stream function:
$\tilde{u} = {\pl\tilde\psi}/{\pl y}$,
$\tilde{v} = -{\pl\tilde\psi}/{\pl x}$.
The resulting perturbation equations (not shown) are
amenable to normal-mode analysis:
\begin{equation}
 (\tilde{\psi}, \tilde{T})(x,\eta,t) = (\tilde{\phi},\tilde{s})(\eta)
     {{\rm e}^{{\rm i}({\alpha}{x}-{\omega}{t})}},
\label{eqn:normalmode}
\end{equation}
where $\alpha$ and $\omega$ are the non-dimensional
wavenumber and frequency, respectively, with $\delta$ and 
$\tau=\delta/U_c$ being the reference length and time scales. 
The amplitudes for the perturbation
stream function and temperature are made dimensionless via
$\phi={{\tilde\phi}/{{U_{c}}\delta}}$ and 
$s = {\tilde{s}}/{T_c}$.
Substituting the normal-mode decomposition (\ref{eqn:normalmode})
into the linearized perturbation equations,
the coupled Orr-Sommerfeld stability equations are obtained:
\begin{eqnarray}
   \left({\phi^{''''}} - 2{\alpha^2}{\phi^{''}} + {\alpha^{4}}{\phi} + {s^{'}}\right) 
   &=& {\rm i}{\alpha}G\left[ \left({f^{'}}-{\omega\over\alpha}\right)({\phi^{''}}
      -{\alpha^2}{\phi}) - {f^{'''}}{\phi} \right] 
\label{eqn_OS1}\\
   {{s^{''}}-{\alpha^2}s} &=& {\rm i}{\alpha}Pr G \left[\left({f^{'}} 
   - {\omega\over\alpha}\right)s  - {h^{'}}{\phi}\right]
\label{eqn_OS2}
\end{eqnarray}
The boundary conditions on $\phi(\eta)$ and $s(\eta)$ are:
\begin{equation}
 \phi(\pm\infty) = \phi^{'}(\pm\infty) = s(\pm\infty) = 0 .
\label{eqn_BC}
\end{equation}

\subsection{Varicose and sinuous modes}

For {\it varicose} modes, both the velocity and temperatures
are {\it symmetric} about the mid-plane ($\eta=0$) which can be
translated into following conditions on $\phi$ and $s$:
\begin{equation}
  \phi(0)=\phi^{''}(0)=s^{'}=0,
 \label{eqn_BCS}
\end{equation}
whereas for {\it sinuous} modes both the velocity and temperatures
are {\it asymmetric} about the mid-plane ($\eta=0$) for which 
the conditions on $\phi$ and $s$ are:
\begin{equation}
  {\phi^{'}}(0)={\phi^{'''}}(0)=s(0)=0.
 \label{eqn_BCA}
\end{equation}
It is  known that the thermal plumes are more unstable to sinuous modes
(Pera \& Gebhart 1971) which has been confirmed in our study too.
Hence, all results are presented only for sinuous perturbations
(\ref{eqn_BCA}).

\subsection{Generalized eigenvalue problem and numerical method}

The linear stability equations (\ref{eqn_OS1}-\ref{eqn_OS2}), 
along with boundary conditions (\ref{eqn_BC}, \ref{eqn_BCA}),
constitute a generalized eigenvalue problem:
\begin{eqnarray}
   {\bf A\Phi} &=& \lambda{\bf B\Phi}.
\label{eqn:evalue1}
\end{eqnarray}
For the {\it temporal} stability analysis, $\lambda=\omega$ is the
eigenvalue, ${\Phi}=(\phi, s)^T$ is the eigenfunction and 
${\bf A}$ and ${\bf B}$ are $2\times 2$ matrix differential operators
whose elements can be easily obtained from (\ref{eqn_OS1}-\ref{eqn_OS2}).
For the {\it spatial} stability analysis, 
the spatial eigenvalue $\lambda=\alpha$ appears nonlinearly in  
(\ref{eqn_OS1}-\ref{eqn_OS2}) which are subsequently transformed into
a linear problem in wavenumber ($\alpha$) by using the `companion-matrix'
method (Bridges \& Morris 1984). 
For this case, ${\Phi}=(\alpha^3\phi, \alpha^2\phi, \alpha\phi, \phi, \alpha s, s)^T$ 
is the eigenfunction and
${\bf A}$ and ${\bf B}$ are $6\times 6$ matrix differential operators;
the non-zero elements of ${\bf A}$ are (with $D={\rm d}/{\rm d}\eta$):
$A_{11}=-{\rm i}Gf'$, $A_{12}= {\rm i}\omega G + 2D^2$, 
$A_{13}={\rm i}Gf'D^2-{\rm i}Gf^{'''}$,
$A_{14}= -D^4 - {\rm i}\omega G D^2$, $A_{16}=-D$,
$A_{53}={\rm i} Pr G h^{'} $, $A_{55}=-{\rm i} Pr G f^{'}$, 
$A_{56}= D^2 + {\rm i}\omega Pr G$,
$A_{21}=1=A_{32}=A_{43}=A_{65}$, and ${\bf B}$ is an unit diagonal operator.

For the temporal stability, the wavenumber, $\alpha$, is real and 
the frequency, $\omega=\omega_r+{\rm i}\omega_i$, is complex, with $\omega_i$ 
being the `temporal' growth/decay rate of the perturbation.
For the {\it spatial} stability, the frequency, $\omega$, is real and 
the wavenumber, $\alpha=\alpha_r+{\rm i}\alpha_i$, is complex, with $\alpha_i$ 
being the `spatial' growth/decay rate.
In either case, the flow is said to be stable/unstable 
if $\omega_i$ or $-\alpha_i >,<0$, respectively,
and neutrally stable if $\omega_i$ or $\alpha_i=0$.

For both temporal and spatial analyses,
the differential eigenvalue problem (\ref{eqn:evalue1})
is transformed into an matrix eigenvalue problem by
discretizating the related differential operators
along the non-periodic $\eta$-direction. 
We have used two numerical methods for discretization (Malik 1990):
(1) the finite difference method with second-order accuracy;
(2) the Chebyshev spectral collocation method.
The resulting matrix-eigenvalue problem 
has been solved by the QZ-algorithm of Matlab.
The growth-rate  and the phase speed of the least-stable mode,
obtained from finite difference and spectral methods, 
were compared for a few test cases with
different number of grid/collocation points ($N=101$ and $151$).
We found that both the growth rate and the phase speed agreed upto 
the third decimal place for two methods. 
Moreover, from a comparison with published literature,
we found that our neutral stability curve for air ($Pr=0.7$)
agrees well with that of Pera \& Gebhart (1971).

\section{Results and discussion}

We have carried out both temporal and spatial stability analyses of a
thermal plume, and most of the results are presented for the temporal case
(except in figure~\ref{fig:Fig3}).

%----------------------
\begin{figure}
\begin{center}
\includegraphics[width=6.00cm]{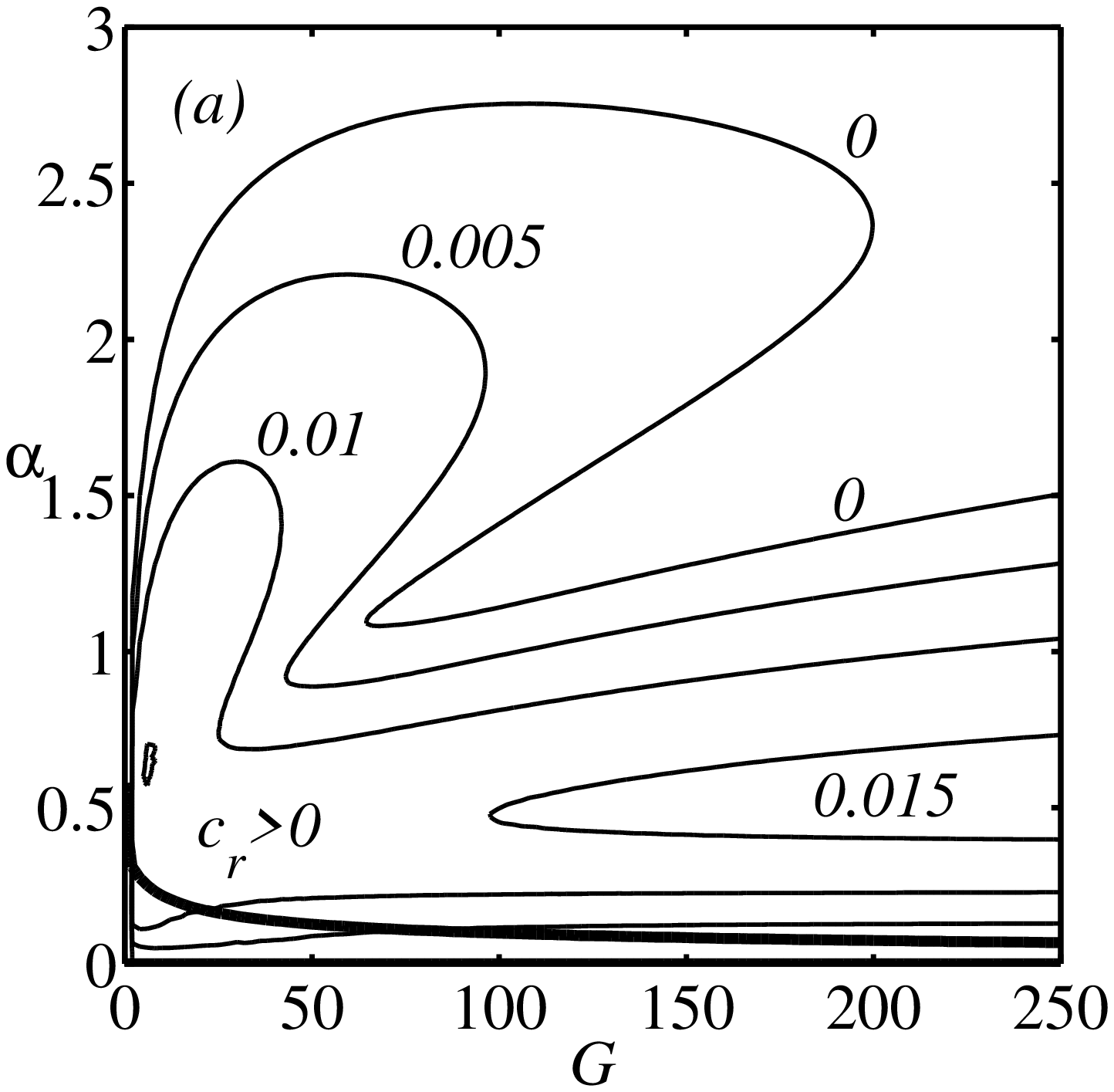}
\includegraphics[width=6.00cm]{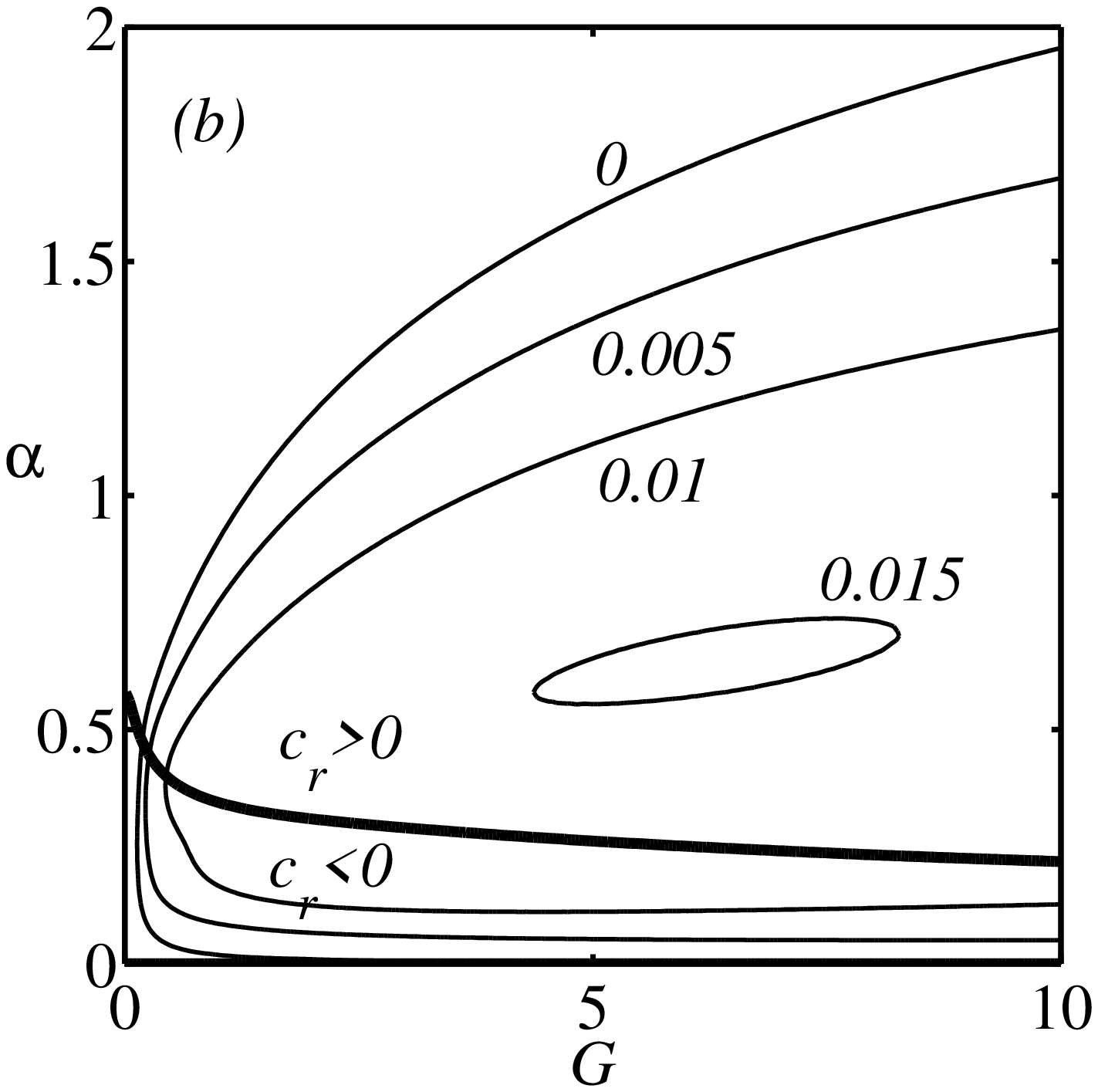}
\end{center}
\caption{
For the {\it temporal} analysis,
the stability diagrams  in the (${\alpha},G$)-plane at $Pr = 200$;
panel $b$ is the zoomed part of low-G region.
}
\label{fig:Fig2}
\end{figure}
%----------------------

%----------------------
\begin{figure}
\begin{center}
\includegraphics[width=6.00cm]{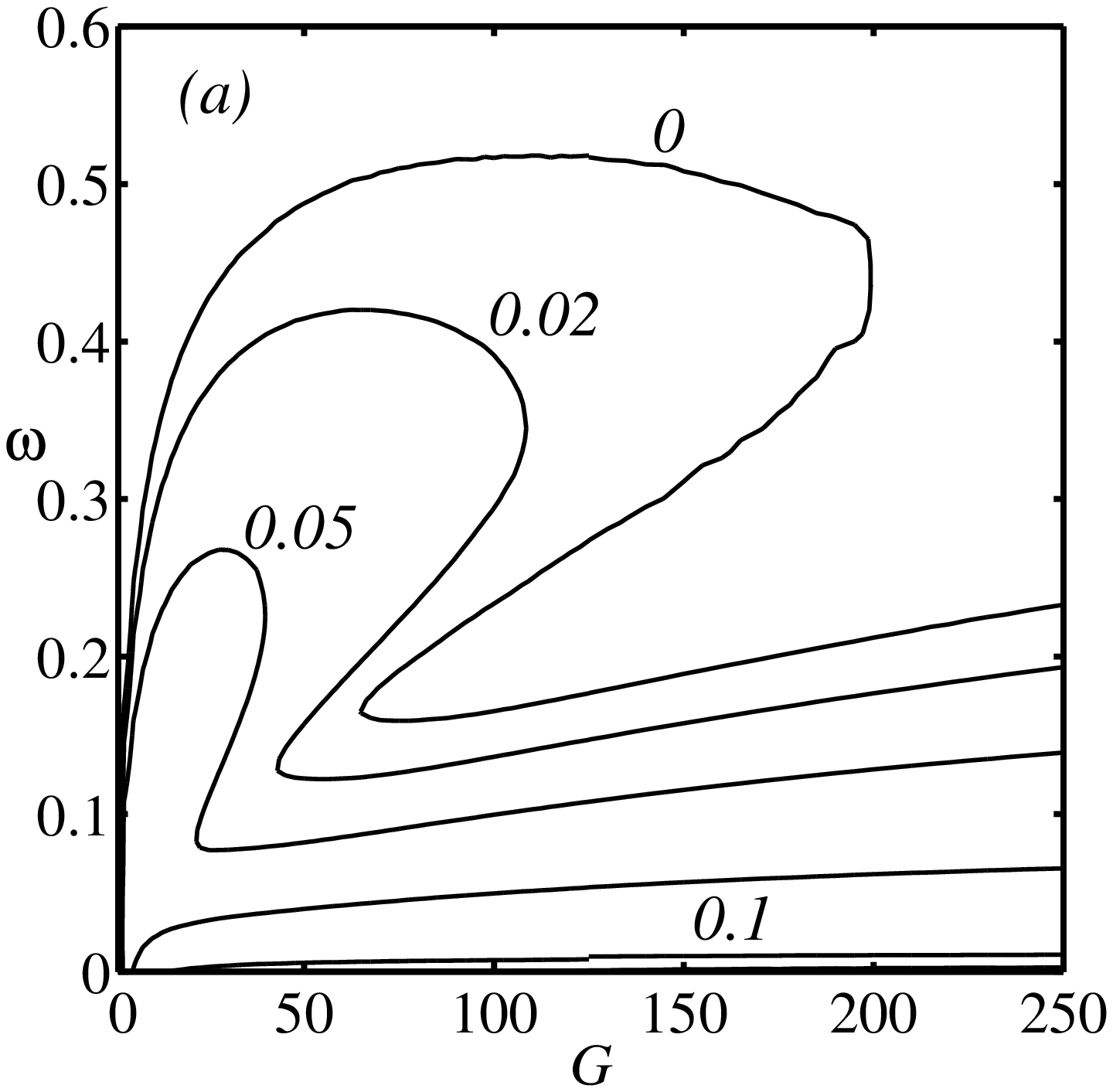}
\includegraphics[width=6.40cm]{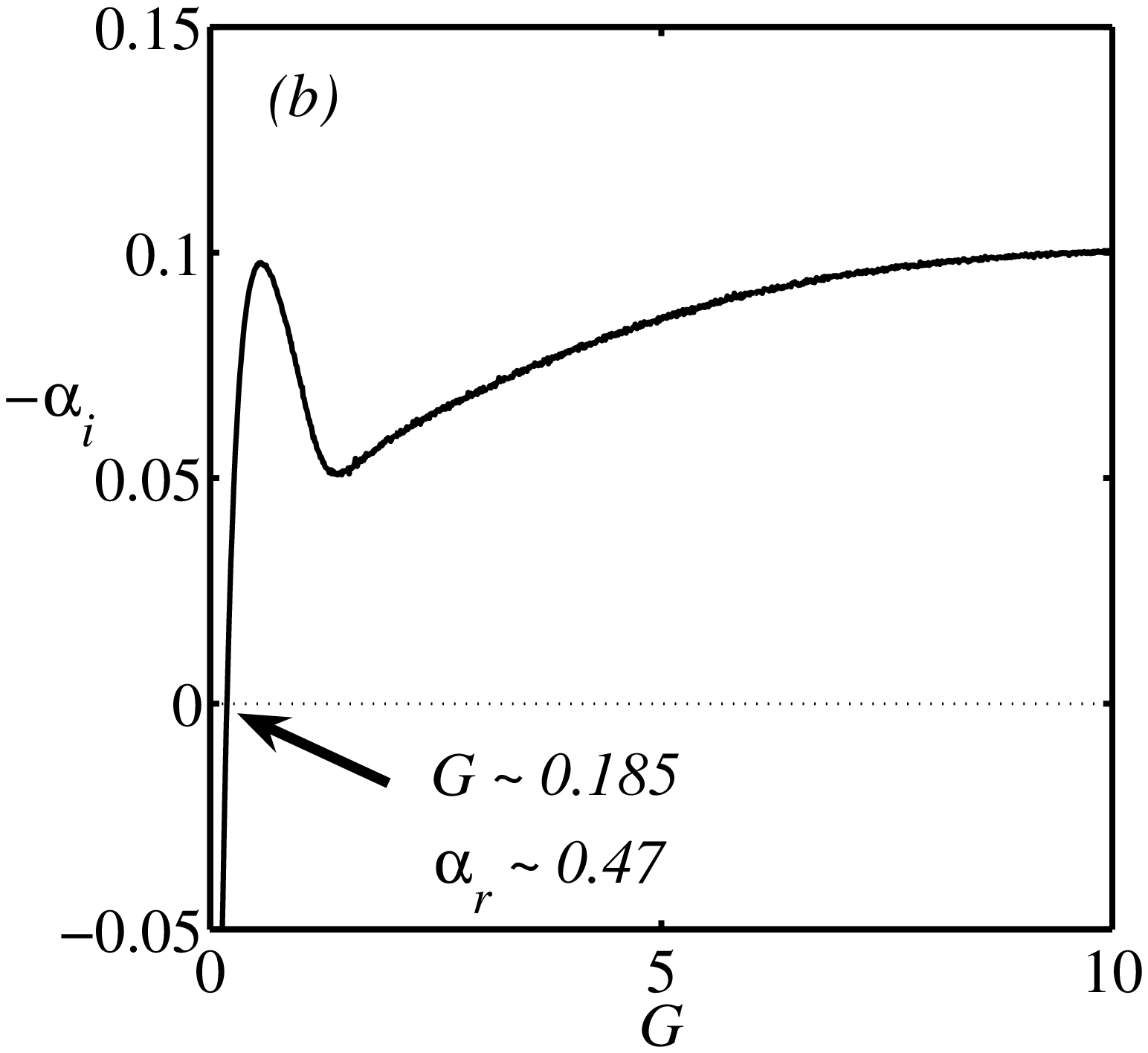}
\end{center}
\caption{
For the {\it spatial} analysis,
($a$) the stability  diagram in the (${\omega},G$)-plane at $Pr = 200$;
($b$) the variation of spatial growth-rate ($-\alpha_i$) with $G$ for $\omega=0$.
}
\label{fig:Fig3}
\end{figure}
%----------------------

\subsection{Results for various Prandtl numbers}

Figure~\ref{fig:Fig2}($a$) displays a typical
stability diagram in the (${\alpha},G$)-plane at high Prandtl numbers ($Pr = 200$)
for the {\it temporal} stability analysis;
figure~\ref{fig:Fig2}($b$) is the zoomed part of the low-G region 
of figure~\ref{fig:Fig2}($a$).
In each panel, the neutral contour ($\omega_i=0$) is  marked by `0', 
and the flow is unstable inside it (see the positive growth rate contours)
and stable outside. 
There are two distinct unstable zones: (a) one at low wavenumbers ($\alpha$) 
that spans the whole range of Grashof number ($G>G_{cr}$), and
(b) the other at relatively higher wavenumbers that spans a limited range of $G$.
Figure \ref{fig:Fig2}($b$) suggests that there is a minimum value of $G$
below which the plume is stable.

The thick line in figure~\ref{fig:Fig2}($a$-$b$) 
demarcates the regions of {\it downstream-propagating}
(phase speed, $c_r=\omega_r/\alpha>0$) and {\it upstream-propagating} ($c_r<0$) 
modes in the ($\alpha, G$)-plane.
The origin of such upstream-propagating modes (at low $\alpha$)
remains unclear to us at present.
We have checked that the locus of $c_r=0$ line in the ($\alpha, G$)-plane
does not change by increasing the size of the computational
domain from $\eta=12$ to $\eta=100$ or by increasing the number of collocation points.
We should point out that the possibility of having upstream-propagating modes
in a plane thermal plume (which exist for any $Pr$ at very small values of $\alpha$)
has not been mentioned in previous works 
(Pera \& Gebhart 1971; Hieber \& Nash 1975; Wakitani 1985). 
Such modes might be analogous to certain backward-propagating modes 
in Ekman boundary layer (Lilly 1966) --
this issue is relegated to a future study.

Figure~\ref{fig:Fig3}($a$) displays the analogue of
figure~\ref{fig:Fig2}($a$) in (${\omega},G$)-plane for
the {\it spatial} stability analysis. 
As expected, the  stability diagram in the (${\omega}, G$)-plane
also contains two unstable loops which are analogues of the two-loops of 
the temporal case, figure~\ref{fig:Fig2}($a$).
Focussing on the zero-frequency modes ($\omega=0$) in figure~\ref{fig:Fig3}($a$),
we plot the variation of the spatial growth rate of the least-stable mode
($-\alpha_i$) with $G$ in figure~\ref{fig:Fig3}($b$).
It is seen that the flow is unstable to $\omega=0$ modes beyond a minimum
Grashof number, $G\sim 0.185$, for $Pr=200$,
and the corresponding real wavenumber is $\alpha_r\sim 0.47$.
This critical point $(G, \alpha_r)=(0.185, 0.47)$ from
the spatial analysis exactly matches
with the intersection point between the neutral curve and the
locus of $c_r=0$ modes in figure~\ref{fig:Fig2}($b$) for the temporal analysis.
This result establishes that the upstream propagating modes ($c_r<0$) for
the temporal case are not an artifact of the numerical method.

%----------------------
\begin{figure}
\begin{center}
\includegraphics[width=6.00cm]{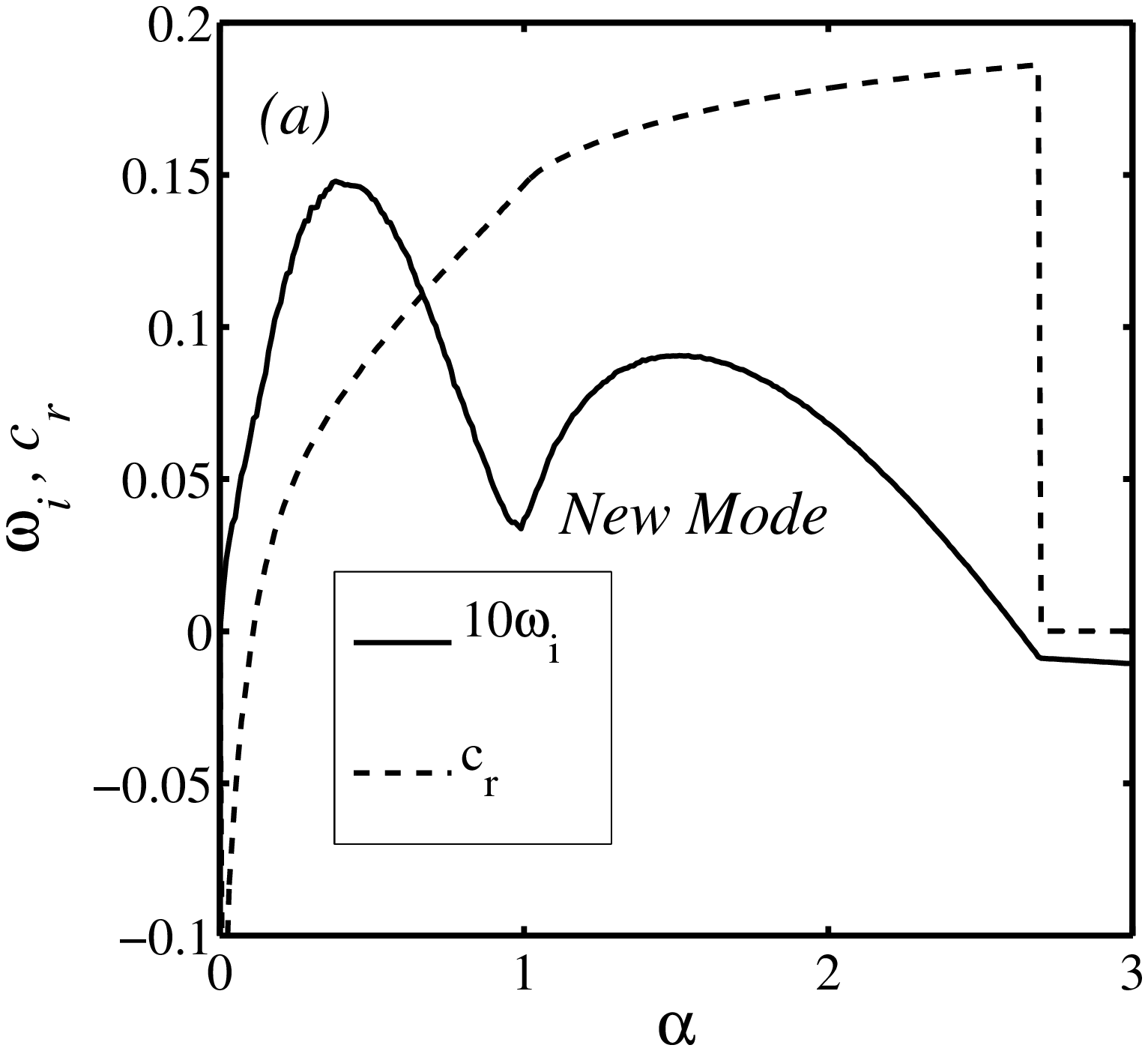}
\includegraphics[width=6.00cm]{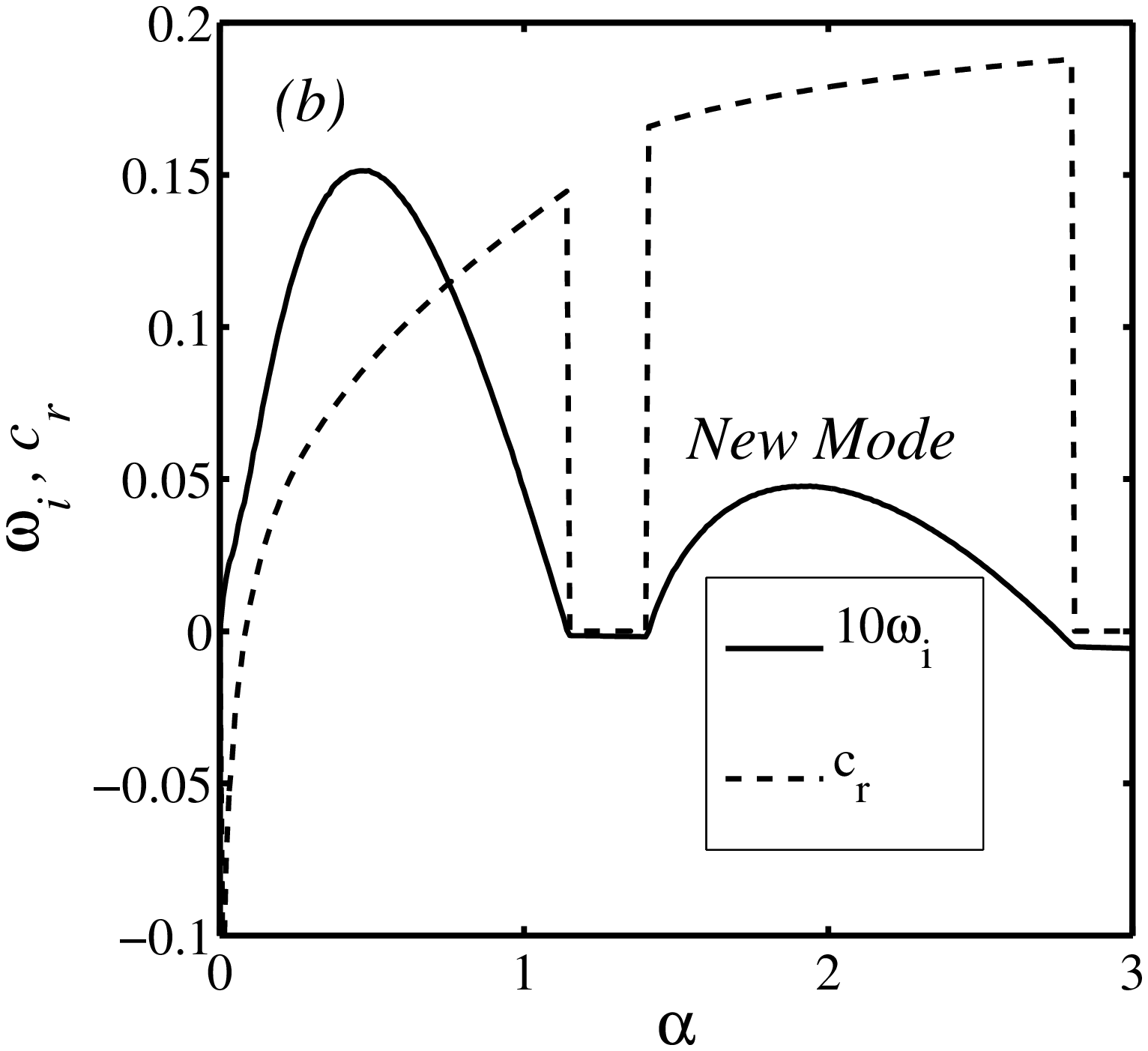}
\end{center}
\caption{
Variations of the `temporal' growth rate ($\omega_i$)
and the phase speed ($c_r=\omega_r/\alpha$) of the least stable mode
with wavenumber $\alpha$ for 
($a$) $G=50$ and  ($b$) $G=100$, with   $Pr=200$. 
}
\label{fig:Fig4}
\end{figure}
%----------------------

Here onwards, we present results only for temporal stability.
Focussing on figure~\ref{fig:Fig2}($a$), 
we show the variations of the growth-rate and  phase speed
of the least-stable mode with wavenumber in 
figures~\ref{fig:Fig4}($a$) and \ref{fig:Fig4}($b$)
for $G=50$ and $G=100$, respectively.
Two humps in each growth-rate curve correspond to two unstable
loops in figure~\ref{fig:Fig2}($a$), and the second hump
is referred to as {\it new mode} since it does
not have an analogue in low-$Pr$ fluids.
The discontinuities in each phase-speed curve  correspond
to crossing of different modes.

\begin{figure}
\begin{center}
\includegraphics[width=6.00cm]{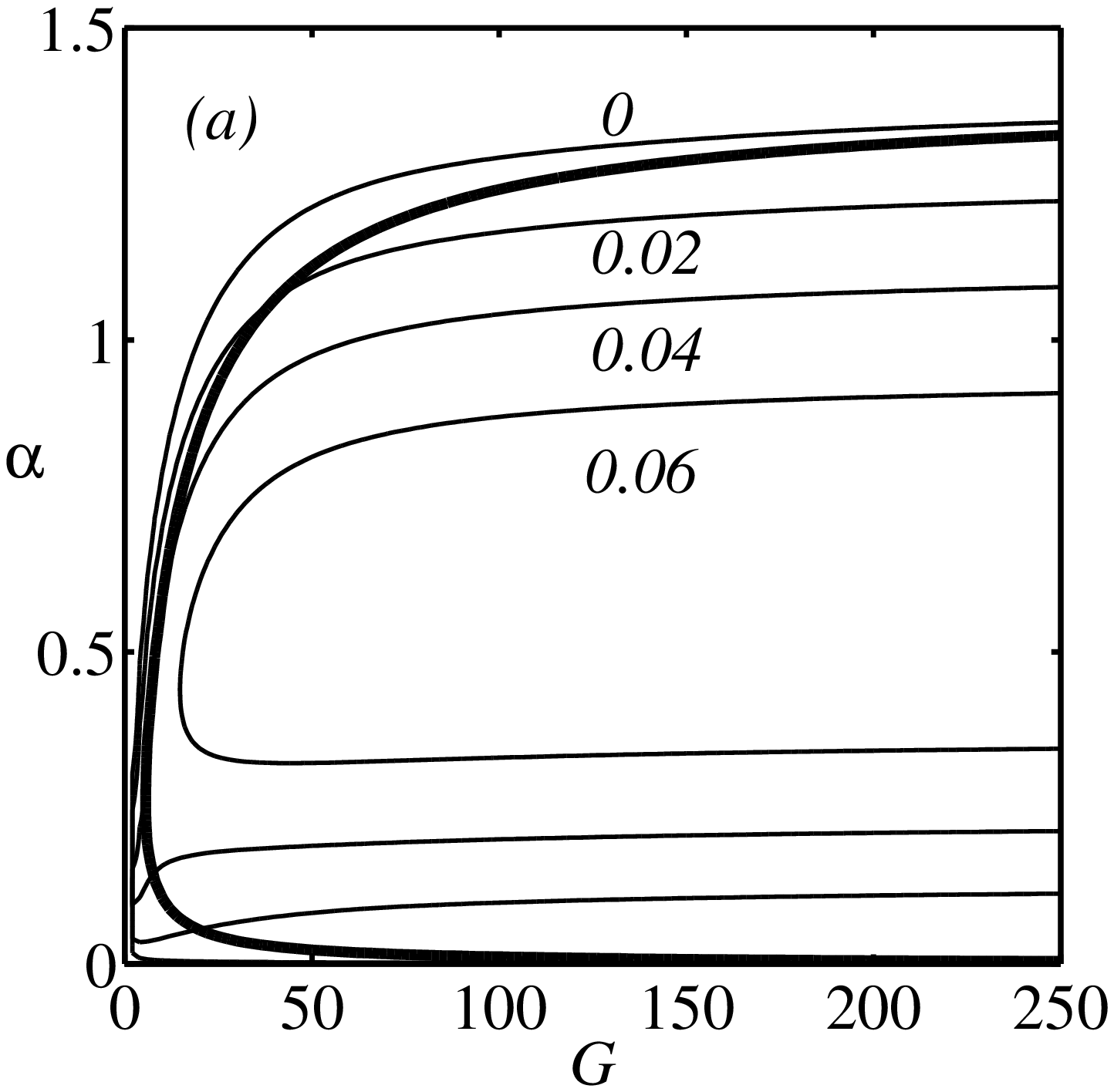}
\includegraphics[width=6.00cm]{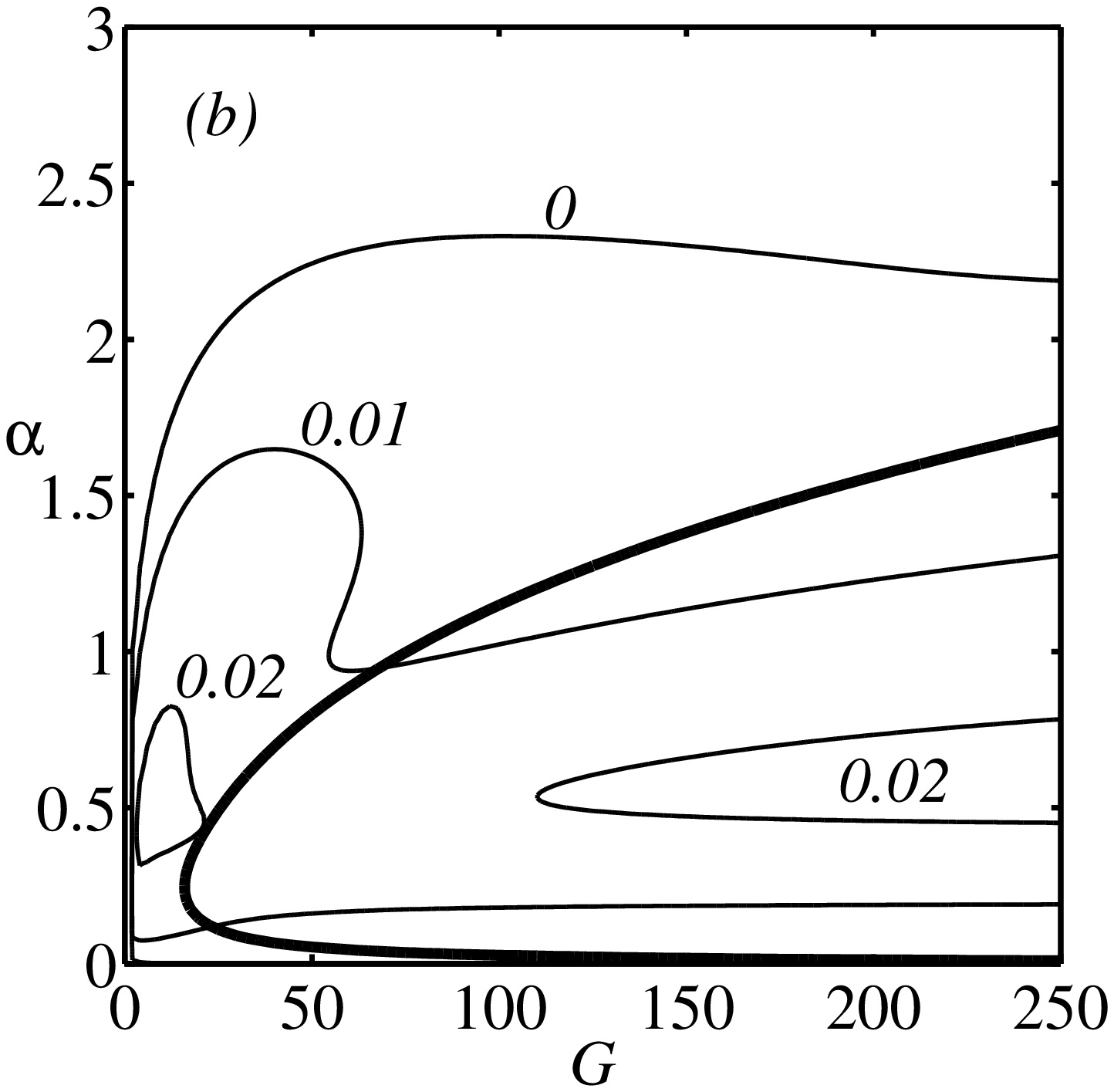}
\includegraphics[width=6.00cm]{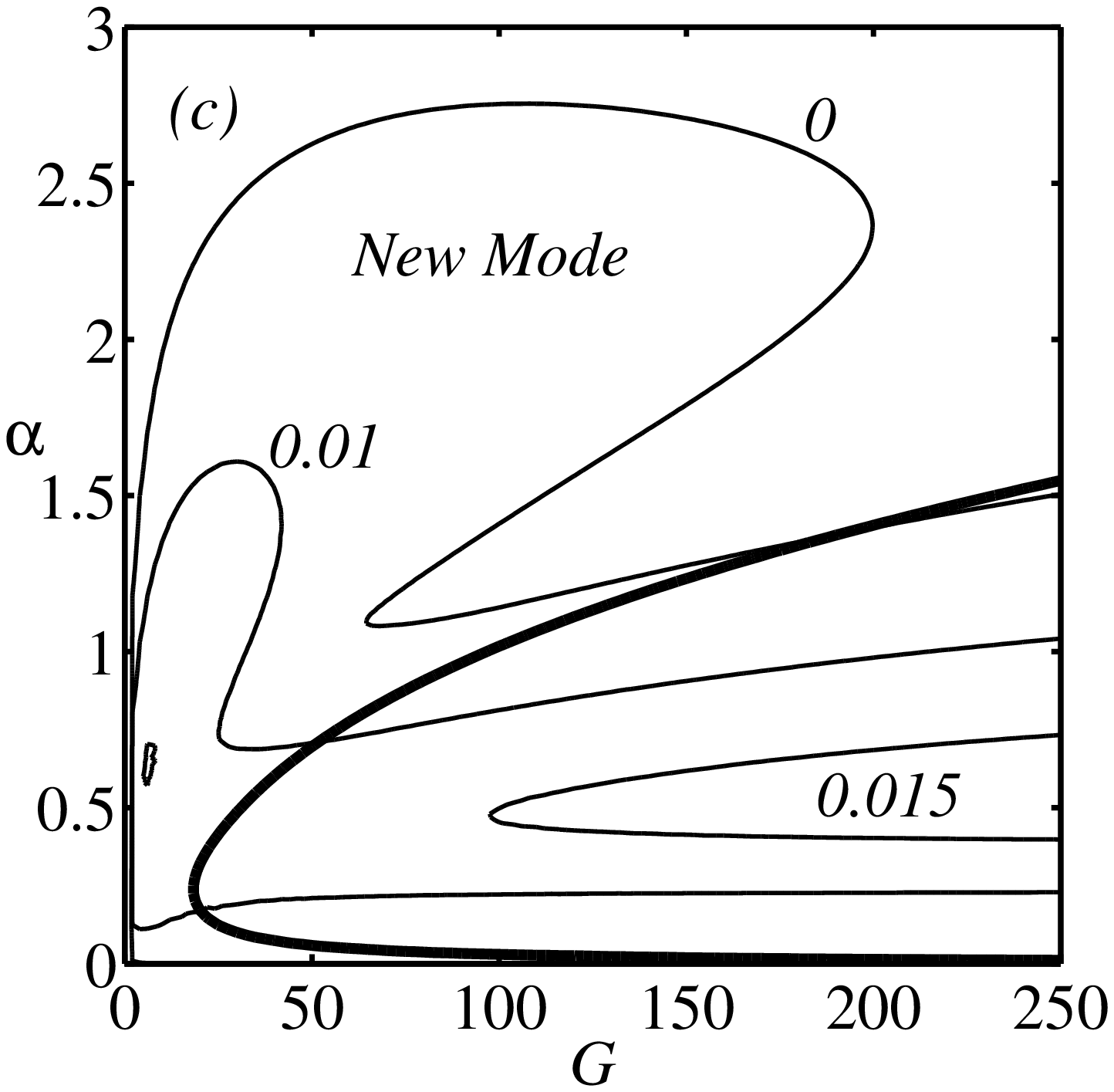}
\includegraphics[width=5.60cm]{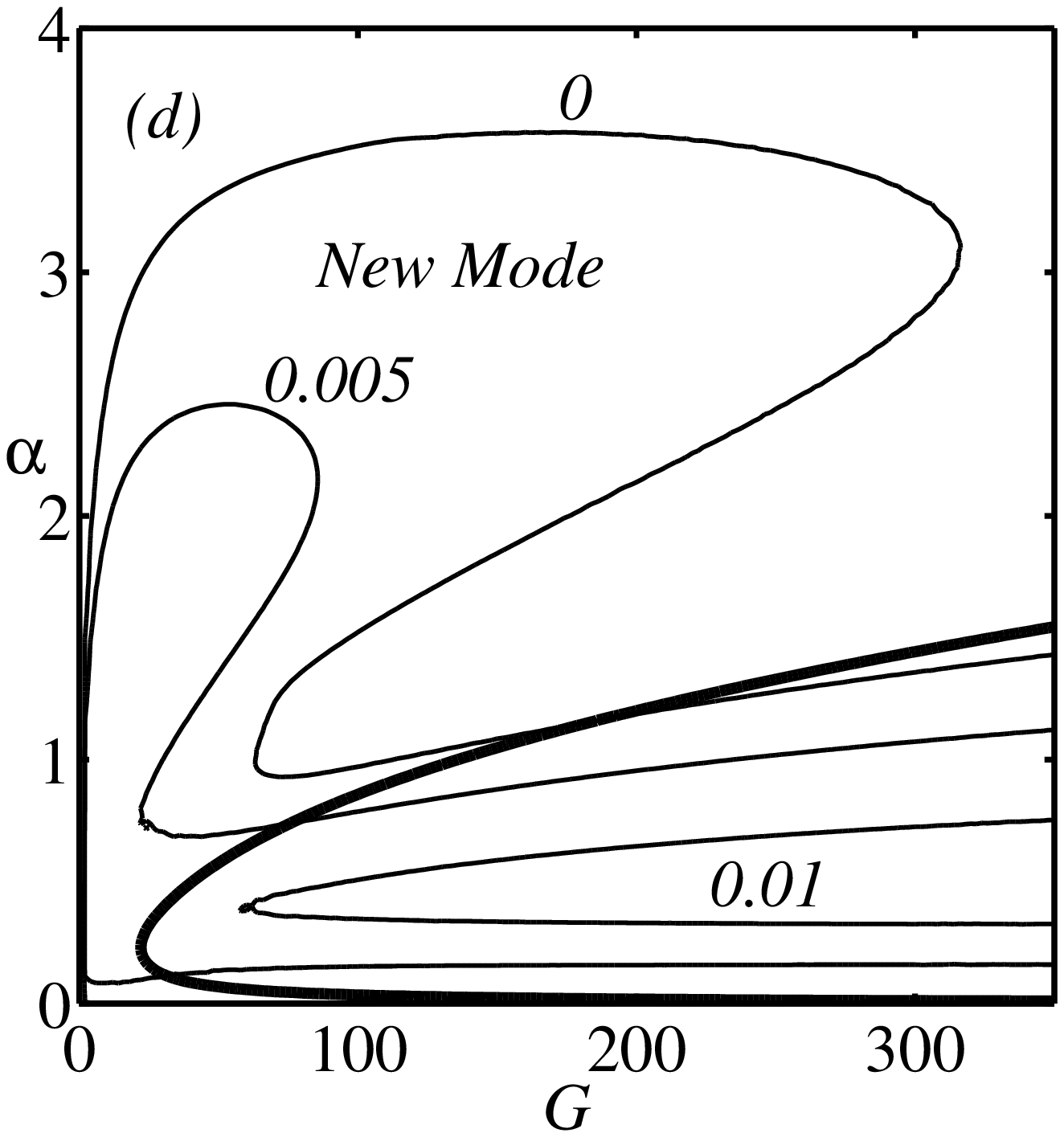}
\end{center}
\caption{Stability diagrams at various Prandtl numbers:
(a) $Pr=0.7$; (b) $Pr = 100$; (c) $Pr =200$; (d) $Pr=500$. 
Thick line in each panel corresponds to the neutral contour for
`uncoupled' stability equations (see \S4.2 for related explanation).
}
\label{fig:Fig5}
\end{figure}

For a range of Prandtl numbers ($Pr = 0.7$, $100$, $200$ and $500$),
the stability diagrams, containing the neutral contour ($\omega_i=0$) 
along with a few positive growth-rate contours ($\omega_i>0$),
are compared in the ($G,\alpha$)-plane in figure~\ref{fig:Fig5}($a$-$d$).
For $Pr=0.7$ (air), the stability diagram has one loop,
and the upper branch of the neutral curve is well 
defined and has an inviscid asymptotic limit: $\alpha=1.3847$ (Pera \& Gebhart 1971). 
In the limit $G\rightarrow \infty$, there exists a 
range of wave-numbers over which the flow is unstable. 
At high Prandtl numbers ($Pr > 100$), as in figure~\ref{fig:Fig5}($c$), 
the neutral curve contains a kink, and there is an additional 
unstable loop at large $\alpha$ and low $G$.
The size of this new unstable loop increases with increasing Prandtl number,
see figure~\ref{fig:Fig5}($d$).
As mentioned before, this new unstable loop is referred to as
a new mode since it does not appear in low-$Pr$ fluids.
Comparing the growth rate contours for different $Pr$ in figure~\ref{fig:Fig5}, 
we find that the growth rate of the least-stable mode decreases with 
increasing Prandtl number, even though the size of the unstable zone
in the $(\alpha, G)$-plane increases in the same limit.
The thick solid contour in each panel of figure~\ref{fig:Fig5}
is explained in the next section.

%----------------------
%----------------------

Now we suggest one possible experiment to realize  this new instability mode.
Suppose a laminar plume is
disturbed by a small-amplitude sinusoidal excitation of the source  
with a specified frequency, $f=\omega/2\pi$ (the temperature at 
the source is constant such that $G=100$, say, in figure 3$a$). 
For small enough $f$ the plume will show
a wavy instability according to the lower instability loop in figure 3($a$),
however, for relatively larger $f$ the  plume is unstable to
our new instability loop
and there is a window of frequencies between two instability modes
over which the plume remains stable. 
It would be interesting to verify this transition scenario at a given Grashof number,
`unstable$\to$stable$\to$unstable' with increasing frequency, 
in experiments of high $Pr$-fluids.

\subsection{Origin of new instability loop: 
coupling of hydrodynamic and thermal fluctuations}

To shed light on the origin of the new instability 
loop at high $Pr$, here we assume that
the velocity and the temperature perturbations are decoupled from each other. 
The full set of stability equations, (\ref{eqn_OS1}-\ref{eqn_OS2}),
can be made independent from each other by dropping $s^{'}$ and ${h^{'}}{\phi}$ 
from equations (\ref{eqn_OS1}) and (\ref{eqn_OS2}), respectively. 
These two sets of equations can now be solved separately to 
determine the least stable eigen-value -- we have verified that
the least stable mode belongs to the Orr-Sommerfeld equation 
(i.e. a purely hydrodynamic mode, eqn. \ref{eqn_OS1}), 
and the energy equation (\ref{eqn_OS2}) always yields a stable mode.

For the uncoupled perturbations, the neutral stability curve
for each $Pr$ is superimposed as a thick solid contour in figure~\ref{fig:Fig5}.
(The flow is unstable inside the thick contour and stable outside.)
A comparison of the thick line in each panel with
the corresponding neutral contour of {\it coupled} stability equations
(denoted by the thin curve $0$) clearly reveals that
the coupling between the hydrodynamic and the thermal disturbance equations
is responsible for the origin of our new instability mode at high $Pr$.
It is observed that the lower parts of the instability loops
in figures~\ref{fig:Fig5}($c$-$d$) closely follow the instability loop of 
the `uncoupled' Orr-Sommerfeld equation, and are, therefore, 
purely hydrodynamic in origin.

It is clear that the coupling terms in the stability equations 
(\ref{eqn_OS1}-\ref{eqn_OS2}) are responsible for appearance of the additional 
instability loop at high Prandtl numbers,
and solving the uncoupled perturbation equations 
would lead to incorrect results.
The importance of this coupling between hydrodynamic and 
thermal perturbations at high $Pr$
can be understood from the fact that  the gradient of the
base-flow temperature, ${h^{'}}$, (which appears
in the energy perturbation equation) increases with Prandtl number:
$h'(\eta) \sim Pr f(\eta)h(\eta) \sim Pr^\epsilon$,
with $0 < \epsilon < 1$, and hence cannot be neglected at large $Pr$.
(From an order-of-magnitude analysis of the pertinent
boundary-layer equations, we find $\epsilon=1/2$.)

\subsection{Analysis of perturbation energy: instability mechanism}

Lastly, to understand the underlying instability mechanism, we analyse
different components of perturbation energy.
The time-evolution equations of perturbation kinetic energy and thermal energy
are  obtained from (\ref{eqn_OS1}-\ref{eqn_OS2}) by 
multiplying them with the corresponding complex conjugate quantity 
$\phi^{\dagger}$ and $s^{\dagger}$,
respectively, and integrating them from $\eta=0$ to $\eta=\infty$. 
Considering the real parts, the resulting evolution equations boil down to
(Nachtsheim 1963; Gill \& Davey 1969)
\begin{eqnarray}
  \frac{{\rm d}{\mathcal E}_K}{{\rm d}t} &\equiv&  
  \omega_i\int_0^\infty e_K {\rm d}\eta \;=\; 
     \int_0^\infty e_{trK} {\rm d}\eta + \int_0^\infty e_{VD} {\rm d}\eta 
      + \int_0^\infty e_B {\rm d}\eta,
\label{eqn:E_K}\\
  \frac{{\rm d}{\mathcal E}_T}{{\rm d}t}  &\equiv&  
  \omega_i\int_0^\infty e_T {\rm d}\eta \;=\; 
    \int_0^\infty e_{trT} {\rm d}\eta + \int_0^\infty e_{TD} {\rm d}\eta ,
\label{eqn:E_T}
\end{eqnarray}
where
\begin{eqnarray*}
   e_K &=& \left({\mid{\phi}\mid^2}+{\alpha^2}{\mid\phi\mid^2}\right), \quad
   e_{trK} =  
     {\alpha}{f^{''}}\left(\phi_r \phi_i^{'}-\phi_r^{'}\phi_i\right), \quad
   e_{VD} = 
     G^{-1}{\mid{{\phi^{''}}-{\alpha^2}\phi}\mid^2}, \label{eq:de37}\\
   e_B &=& 
     G^{-1}\left(s_r\phi_r^{'}+s_i\phi_i^{'}\right), \label{eq:de39}\\
   e_T &=& {\mid{s}\mid^2}, \quad
   e_{trT} \;=\; {\alpha}{h^{'}}\left(\phi_r s_i-\phi_i s_r\right), \quad
   e_{TD} \;=\; 
       - Pr^{-1} G^{-1}\left({\mid{s^{'}}\mid^2}
        +{\alpha^2}{\mid{s}\mid^2}\right), \label{eq:de41}
\end{eqnarray*}
and the suffixes $r$ and $i$ denote the real and imaginary parts, respectively. 
For hydrodynamic fluctuations,
${\rm d}{\mathcal E}_K/{\rm d}t$  represents the rate of change 
of perturbation kinetic energy,
$E_{trK}=\int e_{trK}$ the rate of transfer of kinetic energy from the mean flow 
to perturbations via the Reynolds stress,
$E_{VD}$ the rate of viscous dissipation, 
and $E_B$ the rate of gain of kinetic energy through the {\it buoyancy} force. 
For thermal fluctuations, ${\rm d}{\mathcal E}_T/{\rm d}t$ is the rate 
of change of perturbation thermal energy, 
$E_{trT}$ is the rate of gain of thermal energy from the mean temperature field
and $E_{TD}$ is the rate of dissipation of thermal energy.

%-------------------------
\begin{figure}
\includegraphics[width=5.6cm]{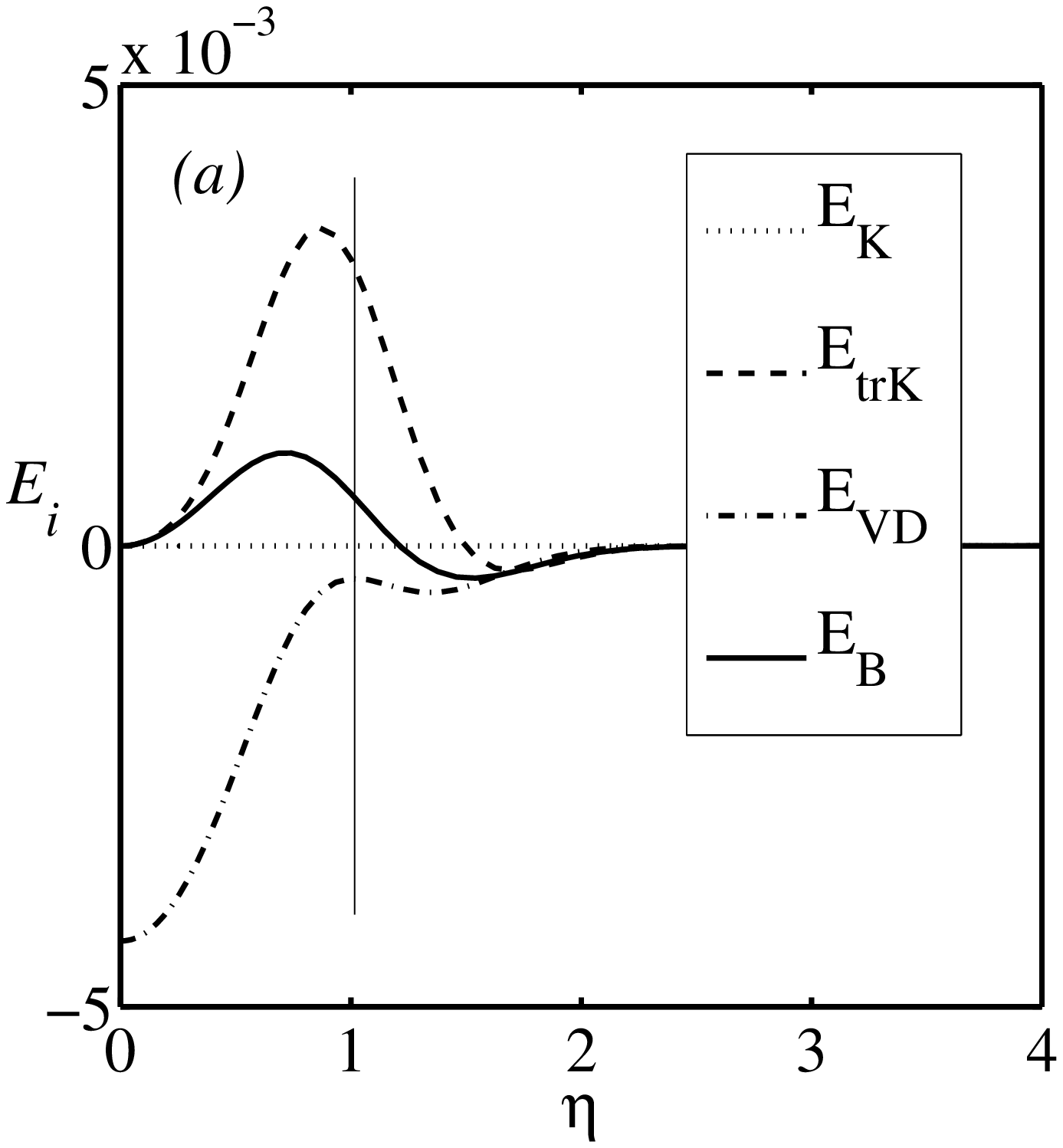}
\includegraphics[width=6.0cm]{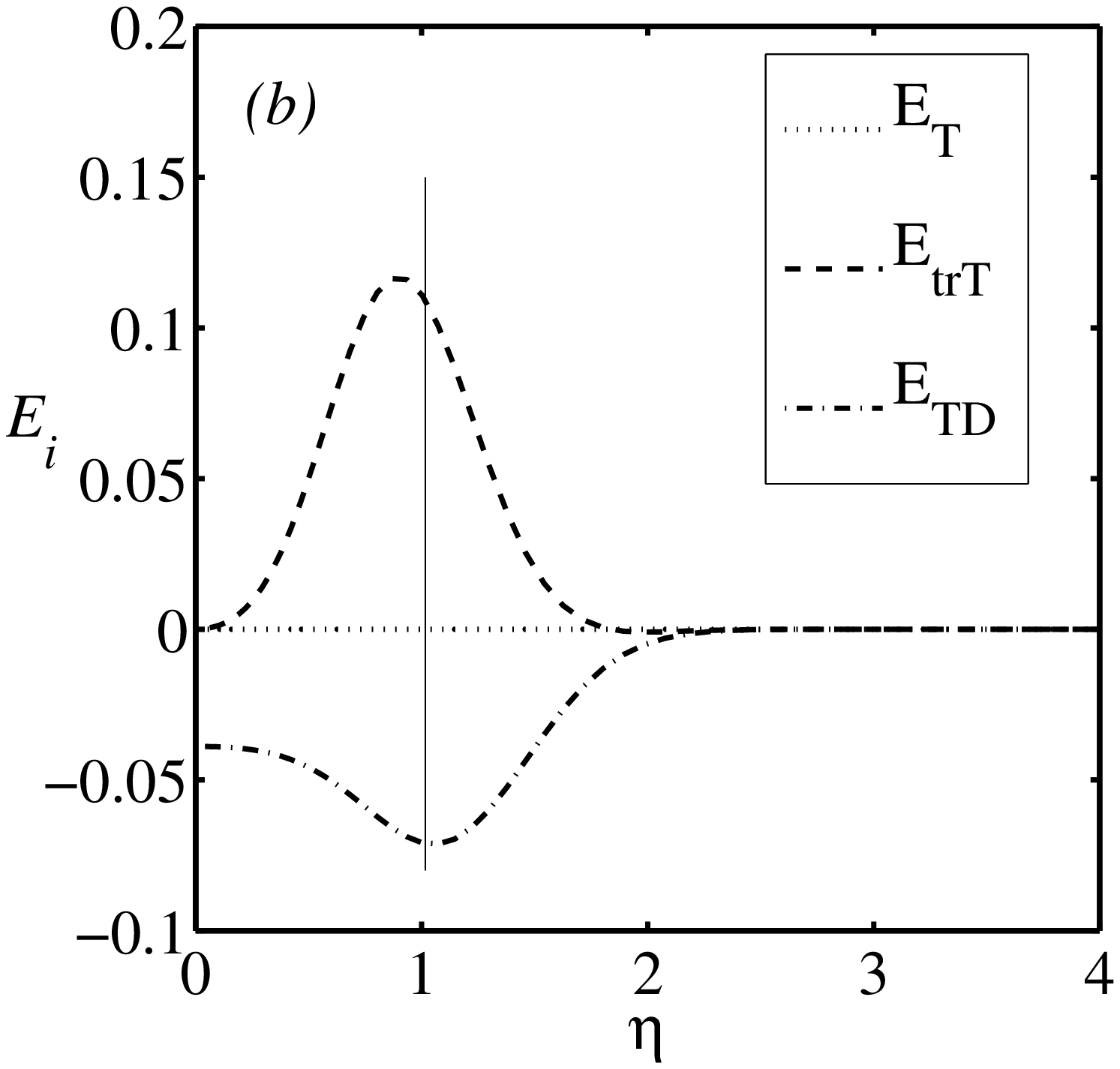}
\caption{Distributions of ($a$) kinetic and ($b$) thermal energies across
the plume width for $Pr=0.7$ (air), $G=100$, $\alpha=1.2923$. 
Vertical line indicates the location of the critical layer.}
\label{fig:Fig6}
\end{figure}
%-------------------------

%-------------------------
\begin{figure}
\includegraphics[width=5.6cm]{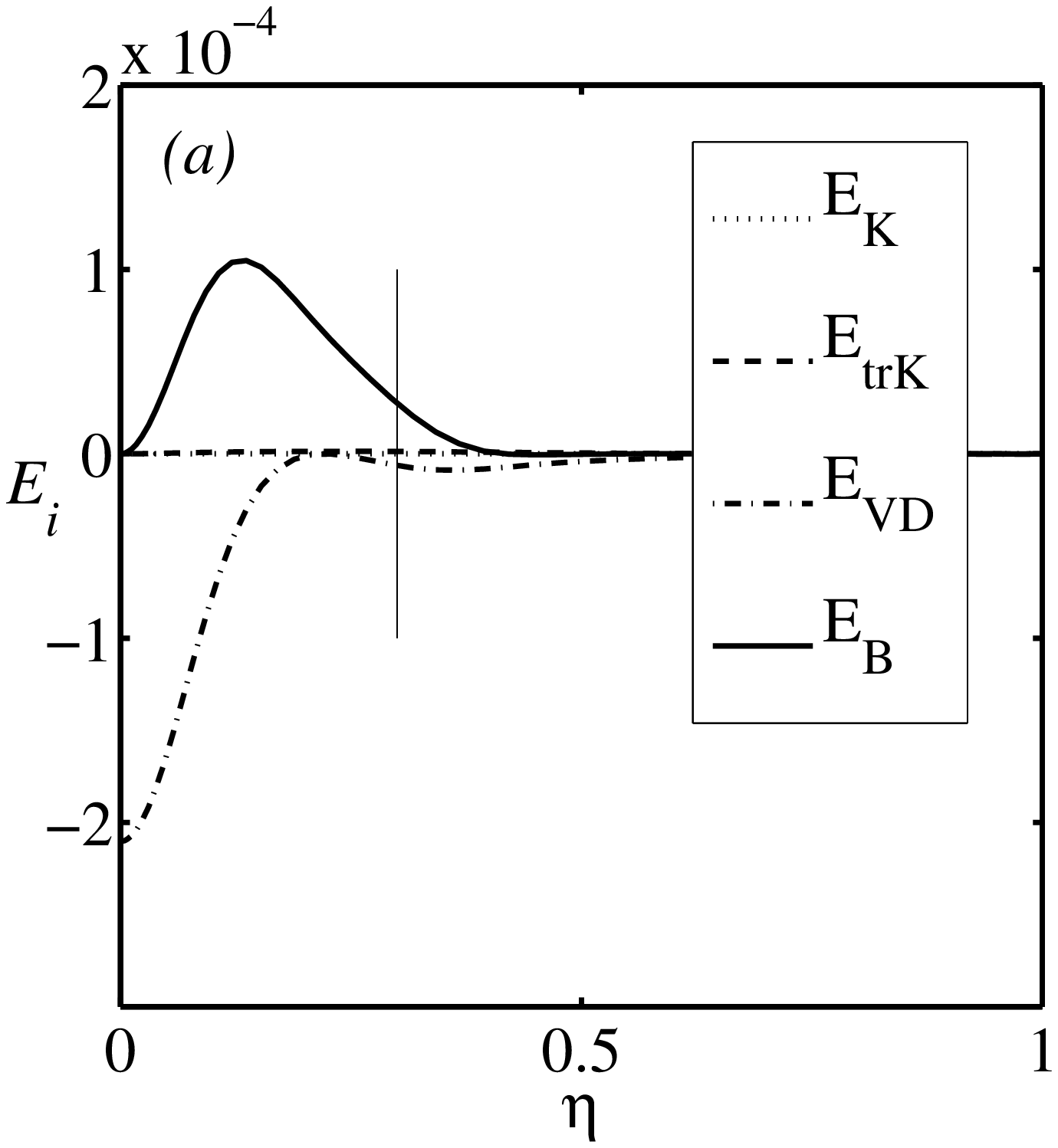}
\includegraphics[width=6.0cm]{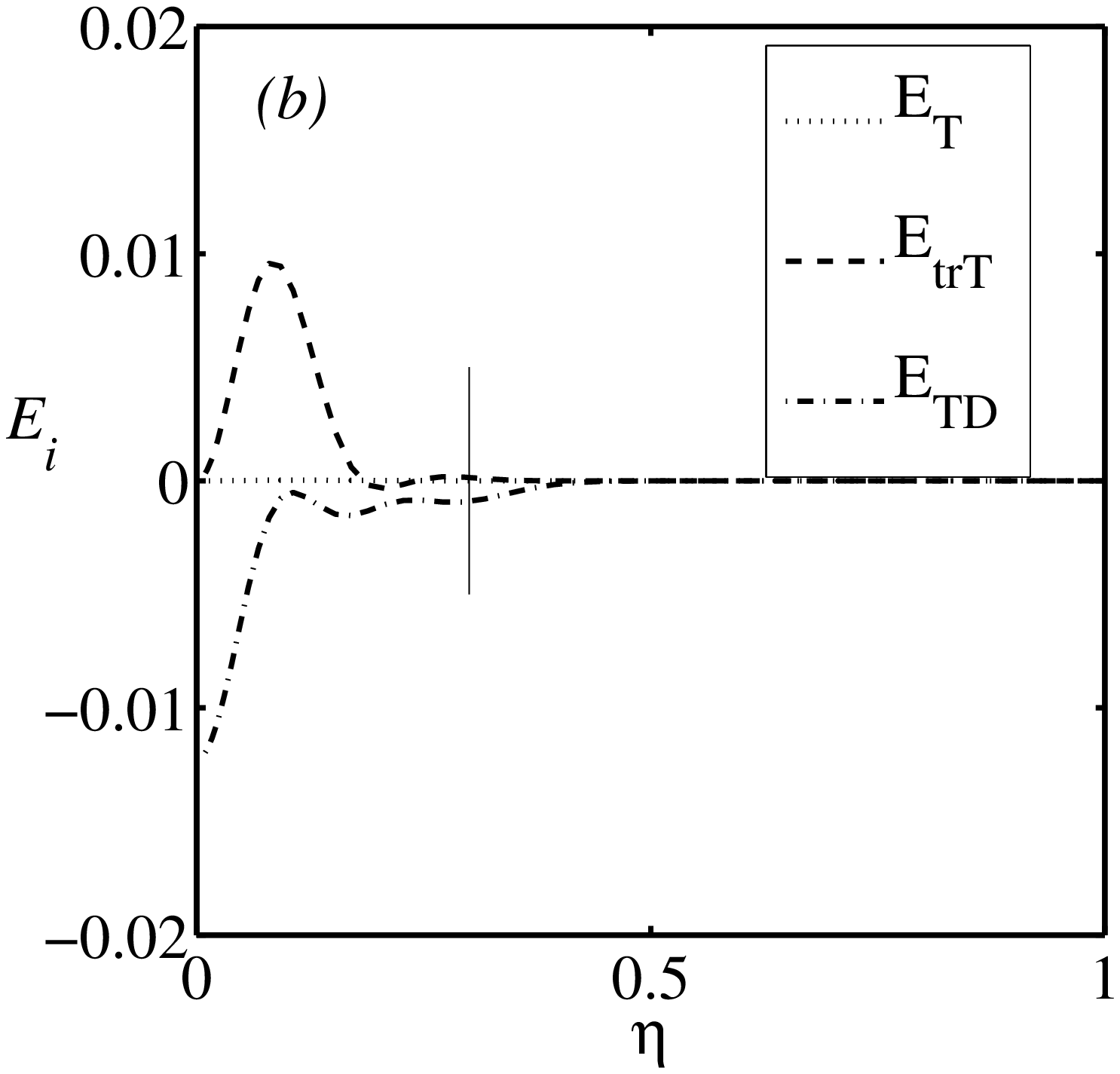}
\caption{
Same as figure 6, but for $Pr=200$, $G=100$, $\alpha=2.74645$.
}
\label{fig:Fig7}
\end{figure}
%-------------------------

The variations of different kinetic and thermal
energy components across the plume width $\eta$ are displayed
in figures~\ref{fig:Fig6} and \ref{fig:Fig7}
for $Pr=0.7$ ($\alpha=1.2923$) and $200$ ($\alpha=2.74645$), 
respectively, at $G=100$. (These two cases correspond to the
neutral modes on the upper branch of the neutral contour
in figures~\ref{fig:Fig5}$a$ and \ref{fig:Fig5}$c$.)
In each figure, the location of the corresponding critical layer is 
indicated by the vertical line.
Since the instability mode is neutral ($\omega_i=0$), 
the net rate of gain of kinetic/thermal energy is 
$E_K=\omega_i\int e_K {\rm d}\eta = 0 =E_T$, denoted 
by the dotted zero-line in each panel.
For $Pr=0.7$, the kinetic energy gained by the perturbation  mainly comes from 
the Reynold's stress term ($E_{trK}$) and 
a small amount is contributed from the perturbation buoyancy force ($E_B$); 
the maximum amount of energy is dissipated at the center line ($\eta=0$) 
by viscous forces ($E_{VD}$). 
With increasing $Pr$, the rate of gain of kinetic energy by Reynolds stress
becomes progressively smaller, and the {\it buoyancy} force ($E_B$) takes
over as the main source of perturbation kinetic energy (see figure~7$a$ at $Pr=200$)
which balances the energy lost due to viscous dissipation.
We conclude that at high $Pr$ the contribution from the Reynold's stress 
term is negligible compared to the gain in kinetic energy by buoyancy force
which drives instability.

\section{Conclusions}

Based on a quasi-parallel stability analysis of a plane thermal plume,
we have uncovered a new instability loop at high Prandtl numbers.
The origin of this new mode is shown to be tied to the coupling between
the hydrodynamic and thermal fluctuations.
The importance of this coupling is tied
to the increasing magnitude of the base-state temperature gradient
with increasing Prandtl number.
It is shown that the perturbation kinetic energy gained 
from the {\it buoyancy} force drives this instability at high $Pr$.
The underlying instability mechanism differs from the well-known
hydrodynamic instability mechanism for which the perturbation
energy is gained from the mean flow via the Reynolds stress.
In future, it would be interesting to analyse the effects of non-parallel
corrections as well as the temperature-dependent
transport coefficients on our new instability loop.

\vspace*{0.5cm}
We acknowledge financial support from two grants
(DRDO/RN/4124 and PC/EMU/MA/35).  
M.A. acknowledges Prof. Vijay Arakeri for motivating this research.

\vspace*{0.5cm}
\begin{center}
{\large\bf REFERENCES}
\end{center}

\begin{itemize}
\item
{Batchelor, G. K.} (1954)
{Heat convection and buoyancy effects in fluids.}
\textit{Q. J. R. Met. Soc.} \textbf{80}, p.~339-359.
\item
{Bridges, T. J. \&  Morris, P. J.} (1984)
{Differential eigenvalue problems in which the parameter appears nonlinearly.}
{\it J. Comp. Phys.} {\bf 55}, p.~437-460.
\item
{Fujii, T.} (1963)
{Theory of the steady laminar natural convection above a horizontal line 
heat source and a point heat source.}
{\it Int. J. Heat Mass Transfer} {\bf 6}, p.~597-606.
\item
{Gebhart, B., Pera, L. \&  Schorr, A.} (1970)
{Steady laminar natural convection plumes above a horizontal line heat source.}
{\it Int. J. Heat Mass Transfer} {\bf 13}, p.~161-171.
\item
{Gebhart,  B., Jaluria, J., Mahajan, R. L. \& Sammakia, B.} (1988)
{\it Buoyancy Induced Flows and Transport.}
Hemisphere Publishing, Washington.
\item
{Gill, A. E. \&  Davey, A.} (1969)
{Instabilities of a buoyancy-driven system.}
{\it J. Fluid Mech.} {\bf 35}, p.~775-798.
\item
{Grossmann, S. \&  Lohse, D.} (2000)
{Scaling in thermal convection: a unifying theory.}
{\it J. Fluid Mech.} {\bf 407}, p.~27-56.
\item
{Hieber, C. A. \& Nash, E. J.} (1975)
{Natural convection above a line heat source: Higher order effects and stability.}
{\it Int. J. Heat Mass Transfer} {\bf 18}, p.~1473-1479.
\item
{Kaminski, E. \&  Jaupart, C.} (2003)
{Laminar starting plumes in high Prandtl number fluids.}
{\it J. Fluid Mech.} {\bf 478}, p.~287-298.
\item
{Lilly, K. L.} (1966),
{On the instability of Ekman boundary flow.}
{\it J. Atmos. Sci.} {\bf 21}, p.~481-494.
\item
{Lister, J. R.} (1987),
{Long-wavelingth instability of a line plume.}
{\it J. Fluid Mech.} {\bf 175}, p.~571-591.
\item
{Majumder, C. A. H., Yuen, D. A. \& Vincent, A. P.} (2004)
{Four dynamical regimes for a starting plume model.}
{\it Phys. Fluids} {\bf 16}, p.~1516-1531.
\item
{Malik, M. R.} (1990)
{Numerical methods for boundary layer stability.}
{\it J. Comp. Phys.} {\bf 86}, p.~376-413.
\item
{Nachtsheim, P. R.} (1963)
{Stability of free convection boundary layers.}
NASA Tech. Note, TN D-2089. 
\item
{Pera, L. \&  Gebhart, B.} (1971)
{On the stability of laminar plumes: Some numerical solutions and experiments.}
{\it Int. J. Heat Mass Transfer}, {\bf 14}, p.~975-984.
\item
{Riley, N.} (1974)
{Free convection from a horizontal line source of heat.}
{\it ZAMP} {\bf 25}, p.~817-828.
\item
{Wakitani, S.} (1985)
{Non-parallel flow instability of a two dimensional buoyant plume.}
{\it J. Fluid Mech.} {\bf 159}, p.~241-258.
\item
{Wang, X.} (2004)
Infinite Prandtl number limit of Rayleigh-Benard convection.
{\it Comm. Pure Appl. Math.} {\bf 57}, p.~1265-1282.
\item
{Worster, M. G.} (1986)
{The axisymmetric laminar plume: Asymptotic solution for large Prandtl number.}
{\it Stud. Appl. Math.} {\bf 75}, p.~139-152.
\end{itemize}
%\end{thebibliography}

\end{document}